# Large Effects of Small Cues:
# Priming Selfish Economic Decisions*

**Avichai Snir****
Department of Economics
Bar-Ilan University
avichai.snir@biu.ac.il

**Dudi Levy**
Department of Economics
Bar-Ilan University
dudilevi1985@gmail.com

**Dian Wang**
Alvarez College of Business
University of Texas at San Antonio
dian.wang@utsa.edu

**Haipeng (Allan) Chen**
Tippie College of Business
University of Iowa
haipeng-chen@uiowa.edu

**Daniel Levy**
Department of Economics
Bar-Ilan University,
Department of Economics
Emory University,
ICEA, ISET at TSU, and RCEA
Daniel.Levy@biu.ac.il

Revised: April 21, 2024

**Keywords:** Self-Selection, Indoctrination, Self-Interest, Market Norms, Social Norms, Economic Man, Rational Choice, Fairness, Experimental Economics, Laboratory Experiments, Priming, Economists vs. Non-Economists

**JEL Codes:** A11, A12, A13, A20, B40, C90, C91, D01, D63, D91, P10

* We thank Raphael Gottweis, Hande Erkut, Rudolf Kerschbamer, Axel Ockenfels, Antonio Cabrales, Martin Rode, Hans Pitlik, and the participants of the First Spanish Public Choice Society Conference at the Universidad de Navarra, the 2024 Innsbruck Winter Summit on (Un)Ethical Behavior in Markets, and the seminar participants at Bar-Ilan University, and ISET-TSU for helpful comments and suggestions. All errors are ours.

** Corresponding author

# Large Effects of Small Cues:
# Priming Selfish Economic Decisions

### *Abstract*

Many experimental studies report that economics students tend to act more selfishly than students of other disciplines, a finding that received widespread public and professional attention. Two main explanations that the existing literature offers for the differences found in the behavior between economists and non-economists are: (i) the selection effect, and (ii) the indoctrination effect. We offer an alternative, novel explanation: we argue that these differences can be explained by differences in the interpretation of the context. We test this hypothesis by conducting two social dilemma experiments in the US and Israel with participants from both economics and non-economics majors. In the experiments, participants face a tradeoff between profit maximization (market norm) and workers' welfare (social norm). We use priming to manipulate the cues that the participants receive before they make their decision. We find that when participants receive cues signaling that the decision has an economic context, both economics and non-economics students tend to maximize profits. When the participants receive cues emphasizing social norms, on the other hand, both economics and non-economics students are less likely to maximize profits. We conclude that some of the differences found between the decisions of economics and non-economics students can be explained by contextual cues.

"The economics students had a much stronger tendency to maximize profits than did the subjects in other groups … even if the economics profession attracts certain types of people, the results still suggest that something is wrong in the way we relate to students in our undergraduate programs."

**Ariel Rubinstein (2006, p. C8)**

## 1. Introduction

In experiments, when participants face a tradeoff between maximizing their profits and adhering to social norms, economics students tend to choose profit-maximizing options. For example, Marwel and Ames (1981) find that compared to students of other disciplines, economics students are more likely to free-ride. Carter and Irons (1991) report that when economics and non-economics students play ultimatum games, economics students tend to offer and accept smaller amounts.

Scholars often interpret these results as suggesting that economists have different personality traits than non-economists. For example, Frank et al. (1996, p. 187) argue that "Economics training… makes [economics students] marginally less likely to cooperate in social dilemmas." Similarly, Rubinstein (2006, p. C9) concludes that "[Economics studies] … contribute to the shaping of a rather unpleasant 'economic man'." Miller (1999, p. 1055) goes even further, arguing that "[Participants] emerge from Economics 101 believing… that not reporting a favorable billing error, in addition to being self-interested, is also the rational and appropriate action to take." Further, differences in personality traits between economists and non-economists might have political effects, because personality traits such as selfishness are correlated with political attitudes (Kerschbamer and Müller, 2020).

The belief that economists have different personality traits is not new (Caplan, 2002). In many policy debates, economists' recommendations are often viewed with great suspicion. Further, economists are often described as heartless and cold disconnected people. For example, a 2017 article in *The Guardian* argues that economic training leads to viewing "human beings as profit-and-loss calculators (and not bearers of grace, or of inalienable rights and duties)".[1] In his January 19, 2024, *Financial Times* column, Tim Hartford concludes that according to the game of Monopoly, economists are not selfish, although taking advantage of mutually beneficial trades in the game leads to accusations

---

[1] Source: www.theguardian.com/news/2017/aug/18/neoliberalism-the-idea-that-changed-the-world, accessed April 18, 2024.

of "ruthless exploitations of innocents."[2]

Consistent with the view that economists have different personality traits, the two most common explanations for the differences in behavior between economists and non-economists are the selection effect and the indoctrination effect (Frey et al., 1993). According to the former, selfish people choose to study economics. According to the latter, economics students are imbued with selfish attitudes during their studies. Both explanations assume that the differences between economics and non-economics students are an expression of stable personality traits, a manifestation of the economic man.[3]

In this paper, we offer a new explanation. We argue that some of the differences between the choices of economists and non-economists can be explained by differences in the interpretation of the context. When participants in an experiment are asked to choose between (a) profit maximization and (b) adherence to a social norm, economics students might interpret the settings differently than other students and, consequently, make choices different from those made by non-economics students.

Our hypothesis is consistent with Akelrof and Kranton (2005, 2008). In their model, people have different norms in different environments. Decisions, therefore, are a function of both the incentives and the environment. Similarly, we argue that even if economics and non-economics students have similar preferences over profit maximization and adherence to social norms, choices depend on the perception of the context. If a subject perceives the situation as a business environment, then s/he would respond by maximizing her profits, following a market norm. If s/he perceives the setting as being a social one, s/he would adhere to the relevant social norm. It is therefore possible that economics students make different choices than non-economics students in experiments because the former are more likely to interpret a situation as business-related than the latter, rather than the possibility that they have different preferences.

---

[2] Source: https://www.ft.com/content/905332d4-a791-483c-a2e6-a80712032112, accessed April 18, 2024. A similar attitude was expressed by a senior Israeli minister who suggested in 2023 that the Governor of the Bank of Israel could be replaced by a robot because his decisions are "disconnected from the people." Source: www.haaretz.com/israel-news/2023-04-03/ty-article/.premium/netanyahus-comms-minister-calls-to-replace-bank-of-israel-head-with-robot/00000187-47b2-dde0-afb7-7fb313e00000, accessed April 18, 2024.
[3] For example, psychologist Carl Rogers described personality as the *self*, organized, permanent, subjectively perceived entity that is at the very heart of all our experiences (Hjelle and Ziegler, 1992, p. 5). Jagelka (2024) discusses the correlations between economic preferences and personality traits and shows that 60% of the variation in the average risk and time preferences can be explained by factors related to cognitive ability and three of the "big five" personality traits.

To test this hypothesis, we conducted two experiments that are based on Rubinstein (2006). In both experiments, participants were asked to play the role of the vice president of a firm that experienced a decline in profits. The vice president must decide how many workers to lay off, effectively choosing between profit maximization (market norm) and retaining workers who have been with a firm for several years (social norm). To manipulate the participants' perception of the context, we used word search puzzles which the participants solved before playing the role of the vice president. In experiment 1, about half of the participants solved a puzzle containing neutral words, such as "toolbox," "umbrella," and "garbage bin" (control treatment). The rest of the participants solved a puzzle containing words such as "inflation," "monopoly," and "income tax" (economics treatment). Consistent with Cipriani et al. (2009) and Brosig et al. (20110), we find evidence in support of the selection hypothesis. More importantly, however, we find that participants subject to the economics treatment, regardless of their field of study, laid off significantly more workers than participants subject to the control treatment. Thus, exposure to words related to economics increased the likelihood that the participants would make profit-maximizing choices.

In experiment 2, we replaced the economics treatment with a communal one, by replacing the economics words/concepts with words related to communal values such as "kindness," "solidarity," and "kindhearted." We find that in the communal treatment, the participants, including economics students, laid off fewer workers than participants in the neutral treatment. Our results, therefore, support the hypothesis that some of the differences between economics and non-economics students can be explained by differences in the interpretation of the context. By increasing the salience of either a business setting (market norm) or a communal setting (social norm), we alter the participants' choices. When the salience of a business (communal) setting is high (low), participants lay off more (fewer) workers.

The paper is organized as follows. In section 2, we review the literature. In section 3, we describe the experimental design. In section 4, we discuss priming and contextual preferences. In section 5, we present the data of experiment 1 and discuss its results. In section 6, we present the data of experiment 2 and discuss its results. In section 7, we address robustness. We conclude in section 8 by summarizing the key findings and suggesting avenues for future research. The online supplementary web appendix contains details of several robustness analyses we performed.

## 2. Literature review

A large body of work, beginning with Marwel and Ames (1981) and Carter and Irons (1991), finds that in experiments, economics students make different choices than non-economics students: economics students tend to make self-interested, profit-maximizing choices while non-economics students tend to make socially minded choices.[4] Rubinstein (2006) asks participants to play the role of a vice president of a firm that faces a drop in profits. Participants are then asked to choose between (a) retaining workers and thereby accepting a drop in profits, or instead (b) laying off the workers to maximize profits. It turns out that economics students (as well as businessmen that studied economics in college), lay off more workers than other students.

These differences between economics and non-economics students seem to persist after graduation. For example, Caplan (2002) and van Dalen (2019) find that years after graduation, economists still hold different beliefs about the world than non-economists. Ambuehl et al. (2023) use a lab experiment to show that politicians' perceptions of social welfare are at odds with standard economic theory.

Frey et al. (1993) offer two possible explanations for the differences between economics students and students of other disciplines: the *selection hypothesis* and the *indoctrination hypothesis*. According to the selection hypothesis, selfish people choose to study economics and, therefore, the differences between economics and other students exist before the students begin their studies. According to the indoctrination hypothesis, training in economics induces students to act selfishly.

Several studies find support for the selection hypothesis, by showing that economics students in their first week of studies make choices that are similar to those of more experienced economics students. For example, Cipriani et al. (2009) and Brosig et al. (2010) replicate Rubinstein's (2006) experiment. They find that economics students in their first week of studies lay off as many workers as more experienced students (Frey and Meier, 2003, 2004; Bauman and Rose, 2011).

---

[4] See Frank et al. (1993), Frank et al. (1996), Selten and Ockenfels (1998), Cadsby and Maynes (1998), Frank and Schulze (2000), Frey and Meier (2003, 2004), Gandal et al. (2005), Kirchgaessner (2005), Rubinstein (2006), Cipriani et al. (2009), Haucap and Just (2010), Bauman and Rose (2011), Wang et al. (2011), Goosens and Méon (2015), Cappelen et al. (2015), and Gerlach (2017).

Support for the indoctrination hypothesis is scarcer. Bauman and Rose (2011) find support for the selection hypothesis, but they also report that non-economics students who have studied at least one economics course were less likely to donate to social programs than other non-economics students. Goosens and Méon (2015) and Laméris et al. (2023) show that compared with other students, economics students are more supportive of trade from the beginning of their studies, supporting the selection hypothesis. They also find that the differences increase with years of study, supporting the indoctrination hypothesis.

A few studies, however, find evidence that is inconsistent with economics students being more self-interested than others. Yezer et al. (1996) find that economics students are at least as likely as other students to return envelopes that contain $10 bills to their owners. Laband and Beil (1999) show that economics professors are not more likely to cheat on their association membership fee payments than professors of other disciplines. Lanteri (2008) finds that the differences between economics and non-economics students in prisoner dilemma games are smaller if the non-economics students know that they are paired with an economics student, suggesting that some of the difference can be explained by differences in expectations. Below, we provide evidence suggesting that some of these differences could also be explained by differences in the contextual cues.

## 3. Experimental design

We ran two experiments. Following Rubinstein (2006), we conducted the experiments in both Israel and the US.

### 3.1. Experiment 1

In Israel, the participants were students at Bar-Ilan University (BIU) and Tel-Aviv University (TAU). In the U.S., the participants were students at the University of Texas at San Antonio (UTSA).

At BIU and TAU, we conducted the experiment during November 2015–June 2016 by entering classes 15 minutes before the end of the lesson and asking the students to take part in an experiment. At UTSA, the experiment was conducted in November 2019, in a Behavioral Laboratory of the university.

At the start of the experiment, the participants were given a questionnaire composed of three parts. The first part was a word search puzzle, which contained 15 words that the participants had to find. They were given 5 minutes for this task, and they were asked to find as many words as they could. In Israel (the U.S.), the participants were awarded NIS 1 ($ 0.25) for every word that they found.[5]

We had two types of puzzles, and the participants were randomly assigned to one of them. Participants in the *control* treatment received a puzzle containing neutral words, such as "toolbox," "umbrella," and "garbage bin." Participants in the *economics* treatment received a puzzle containing words related to economics, such as "inflation," "monopoly," and "income tax." The words in the two puzzles were chosen such that the total length of the words would be similar, to ensure that the puzzles were of similar difficulty levels.[6]

After solving the puzzles, the participants proceeded to the second part of the experiment, which contained the social dilemma question from Rubinstein (2006):

"Assume that you are the vice president of RPG company. The company provides extermination services and employs administrative workers who cannot be fired and 196 non-permanent workers who do the actual extermination work and can be fired. The company was founded 5 years ago and is owned by three families. The work requires only a low level of skills, so each worker requires only one week of training. All the company's workers have been with the company for three to five years. The company pays its workers more than the minimum wage. A worker's wage, which includes overtime, amounts to about NIS 7,000 per month.[7] The company provides its workers with all the benefits required by law.

Until recently, the company was very profitable. As a result of the continuing recession, however, there has been a significant drop in profits though the company is still in the black. You will soon be attending a meeting of the management at which a decision will be made as to how many workers to lay off. RPG's Finance Department has prepared the following forecast for annual profits:

---

[5] At the time we conducted the experiment, the average NIS–US$ exchange rate was NIS 3.56 for $1.

[6] For a copy of the questionnaire, see Appendix A.

[7] In Rubinstein's original study, the workers' monthly wage was set at NIS 4,500, on average. We raised it to NIS 7,000 because at the time Rubinstein ran his experiments, the minimum wage was NIS 3,335, while during the period in which we conducted our experiments, the minimum had risen to NIS 5,300.

| Number of Workers who Will Continue to be Employed | Expected Annual Profit in NIS Millions |
|---|---|
| 0 | Loss of 8 |
| 50 | Profit of 1 |
| 65 | Profit of 1.5 |
| 100 | Profit of 2 |
| 144 | Profit of 1.6 |
| 170 | Profit of 1 |
| 196 (no layoffs) | Profit of 0.4 |

I will recommend continuing to employ □ 0 □ 50 □ 65 □ 100 □ 144 □ 170 □ 196 of the 196 workers in the company."

After answering this question, the participants filled out the final part of the questionnaire, which contained socio-demographic questions. Upon completion of the questionnaire, we paid the participants, and the experiment was over.

### 3.2. Experiment 2

In Israel, we run the experiment from October 2020–November 2020. We could not run the experiment in person because of the COVID-19 lockdowns. Instead, we sent a link to the webpage containing the experiment to students of Bar-Ilan University, the Hebrew University, the Open University, Tel-Aviv University, Netanya Academic College, the College of Management Academic Studies, and Lev Academic Center.

On the experiment webpage, the participants were randomly assigned to one of two groups, i.e., control treatment, and communal treatment. Next, the participants were shown a list of 15 words and were given 1 minute to memorize them. Then, they were asked to write all the words that they could recall. Participants who were assigned to the control treatment were shown the same neutral words as in the control treatment of experiment 1. In the communal treatment, they were shown words related to communal values, such as "equality," "charity," and "social norm." As in experiment 1, we chose words such that the total length of the words in the treatments would be similar.

After listing the words that they could recall, the participants answered the Rubinstein (2006) question, followed by the same socio-demographic questions as in experiment 1. Participants were not paid for their participation.

In the U.S., the experiment was conducted in April 2023 in person in a laboratory,

after the pandemic-related restrictions were lifted. The experiment was conducted following the same protocol as we did in experiment 1. The only difference was that we replaced the economics treatment with a communal treatment, where the participants were assigned a word search puzzle that contained 15 words related to communal values.

## 4. Priming and contextual preferences

In Akelrof and Kranton (2005, 2008), people follow different norms in different settings. For example, a person might approach a stall at a bazaar and haggle with the seller, even if this person would never consider haggling with the seller in some other setting, say at a shopping mall, for example.

Similarly, a person that perceives a situation as set in a business environment, is likely to seek profit maximization. The same person might make a different decision if s/he were to perceive the situation as a cocktail party at the boss' office.

Research in psychology suggests that the choice of relevant norms depends on the interpretation of the context and is often affected by the cues. As Smith et al. (2003, p. G-12) note, cues received before exposure to a situation might lead to *priming*, the increased accessibility or retrievability of information stored in memory. Priming is particularly important in ambiguous contexts, i.e., situations that can be interpreted in more than one way. That is because when the context is ambiguous, a cue received before the situation might determine which parts of the environment receive greater attention and, therefore, how the situation is interpreted.

Previous research has shown that priming can affect economic decisions (Kay et al., 2004; Vohs et al., 2006; Wang et al., 2011; Cohn et al., 2014, Bansal et al., 2016).[8] For example, Liberman et al. (2004) find that participants who played a prisoners' dilemma game that was titled the "Wall Street Game" were significantly more likely to defect than participants who played the same game but with the title, the "Community Game."[9]

We believe that the dilemma used in Rubinstein (2006) is an example of an ambiguous situation in which participants' decisions might depend on the cues that they receive before the experiment. A participant who focuses on the firm's dwindling profits is more likely to respond by making a profit-maximizing decision. In comparison, a participant focusing on the workers who face unemployment is more likely to consider their welfare before making the decision.

---

[8] Van Oers et al. (2005) find evidence of context-dependent behavior in birds.

[9] An alternative model for understanding priming is *top-down thinking* (also known as *schematic processing*). According to top-down thinking, memory is organized in structures known as *schemas*. A *schema* is an organized set of beliefs and knowledge about people, objects, events, and situations. Top-down thinking is the process of searching in memory for the schema that is most consistent with the incoming data. Schemas are useful because they permit us to organize enormous amounts of data very efficiently. For example, top-down thinking allows us to readily categorize consumables as either food or drink and then put one on a plate and the other in a glass (Smith et al., p. 646).

This suggests that some of the differences found in Rubinstein's (2006) experiment between the decisions made by students majoring in economics and other students, were perhaps driven by differences in the interpretation of the context. Economics students are more likely to pay attention to the profit-maximization aspects of the dilemma because they discuss economic decisions regularly as part of their academic training. Non-economics major students, on the other hand, are less exposed to markets and market norms in their academic training, and therefore, are more likely to think about the workers facing layoffs.

We test this hypothesis in experiments 1 and 2. In experiment 1, we randomly prime half of the participants with words related to the world of economics. We hypothesize that the exposure will heighten the salience of the business aspect of the problem, increasing participants' likelihood to lay off a larger number of workers.

In experiment 2, we randomly prime half of the participants with words related to communal values. We hypothesize that participants who are primed with communal values will pay greater attention to the workers facing unemployment. We, therefore, hypothesize that exposure to words related to communal values will lead participants to lay off fewer workers than those in the control group.

## 5. Experiment 1: data and results

### 5.1. Data

Panel 1 of Table 1 presents a summary statistics of the participants in experiment 1.[10] In Israel (the US), we had 544 (99) participants. The average age of the participants was 23.3 (22.3). 65.6% (40.4%) of the participants studied economics,[11] 59.9% (37.4%) were women, and 10.3% (7.1%) were married.

---

[10] Summary statistics of the participants by treatment for both experiments are given in Appendix E.
[11] We define a participant as an economics student if s/he studies economics, accounting, business administration, banking and finance, or management.

In Israel, 52.8% of the participants took part in the experiment during their first week of studies. 10.8% of the Israeli participants reported that they vote for center-left or left parties. In the U.S., 40.4% of the participants reported that they vote for the Democratic party. Israeli (U.S.) participants retained, on average, 133.8 (153.7) of the 196 workers.[12]

### 5.2. Results

Figure 1 depicts the average number of workers that the participants retained in each treatment. The vertical lines indicate the standard deviations of the means. In both Israel and the U.S., participants in the economic treatment retained, about 10 workers less on average, than participants in the control treatment. The differences are statistically significant. The *t*-statistics are 3.81 for Israel ($p < 0.01$), and 2.25 for the US ($p < 0.03$). See Appendix D for more details about the distribution of the participants' responses.

To assess whether this result is driven by non-economics students, Figure 2 depicts the average number of workers retained by economics students and by other students, separately. Panel A gives the results for Israel, and Panel B for the U.S. In both countries, both economics and non-economics students who were included in the economic treatment retained fewer workers than participants in the control treatment. Therefore, it seems that the differences in the number of workers retained are not driven by non-economics students. Rather, they are common across all the participants.

As a formal test, we estimate a series of OLS regressions. In the regressions, the dependent variable is the number of workers retained. In the baseline model, the independent variables are a dummy for the economic treatment, which equals 1 if the participant took part in the economics treatment and 0 otherwise, and a dummy for economics students, which equals 1 if the participant studied economics and 0 otherwise. We report robust standard errors, clustered by sessions. We report the results in Table 2.

Column 1 reports the results for Israel and column 4 for the U.S. We find that in Israel (the US) participants in the economic treatment retained 10.69, $p < 0.01$ (10.27, $p < 0.05$) fewer workers than participants in the control treatment. Therefore, in both countries, exposure to economic cues leads to a significant decrease in the number of workers retained. We also find that consistent with previous studies (Rubinstein, 2006, Cipriani et al., 2009) economics students retain fewer workers than other students. In Israel, they retained 13.18 ($p < 0.01$) fewer workers. In the U.S., they retained 7.09 fewer workers, but the coefficient is not statistically significant.

---

[12] The differences between Israeli and US participants are consistent with Roth et al. (1981) who find that Israeli participants offer and accept smaller sums in ultimatum games than US participants. It is also consistent with Creedy et al. (1999) who find that Israeli participants are less inequality averse than Australian participants.

In columns 2 (for Israel) and 5 (for the U.S.), we add further controls: Woman – a dummy that equals 1 if the participant is a woman, and 0 otherwise, married – a dummy that equals 1 if the participant is married, and 0 otherwise, voting left-wing/democrats – a dummy that equals 1 if an Israeli (the U.S.) participant responded that s/he votes for left or center-left parties (votes for the Democratic party), and 0 otherwise, religious – a dummy that equals 1 if the participant identified herself as religious or ultra-religious, and 0 otherwise, employment – a dummy that equals 1 if the participant works either part-time or full time, and 0 otherwise, academic – a dummy that equals 1 if both parents of the participant have academic degrees, and 0 otherwise, the participant's age in years, and the number of words that the participants found in the puzzle.

We find that adding these variables has little effect on the main result. The economic treatment coefficients remain almost unchanged and statistically significant.

In columns 3 and 6, we add an interaction term between the economic treatment and economics students. The coefficient of this variable shows whether economics students are affected by economic treatment differently than other students. In Israel (column 3), we find that the main effect of the economic treatment, −12.47, remains statistically significant at the 1% level. In the U.S. (column 6), the size of the main effect of the economic treatment remains almost unchanged relative to the previous columns, −10.37, and is marginally significant. In both countries, the interaction terms' coefficients are much smaller than the main effects and are not statistically significant.

It, therefore, seems that the effect of the economic treatment is similar across economics and non-economics students. In both Israel and the U.S., priming participants by exposing them to words related to economics leads them to retain fewer workers.

## 6. Experiment 2: Data and results

### 6.1. Data

Panel B of Table 1 presents summary statistics of the participants in experiment 2. In Israel (the U.S.), we had 383 participants (212), with an average age of 30.1 (21.9). 59.0% (44.8%) of the participants studied economics, 53.0% (46.2%) were women, and 36.6% (6.1%) were married. In addition, 13.3% (30.2%) of the Israeli (U.S.) participants reported that they vote for center-left or left parties (the Democratic party). We also find that Israeli (U.S.) participants retained, on average, 144.3 (152.6) of the 196 workers.

### 6.2. Results

Figure 3 depicts the average number of workers that the participants retained in each treatment. The vertical lines indicate the standard deviations of the means. In Israel (the US), participants in the communal treatment retained on average 11.9 (8.2) workers more than the participants in the control treatment. The differences are statistically significant: the $t$-statistics are 3.57 for Israel ($p < 0.01$), and 2.37 for the U.S. ($p < 0.02$).

To test whether priming by communal values affects economists as well as other students, Figure 4 depicts the average number of workers retained by economics students and by other students, separately. When we focus on the control groups, in this data we do not find significant differences between economics and non-economics students. In Israel (the U.S.), the $t$-statistics for testing the null hypothesis of no differences between the average number of workers retained by economics vs. non-economics students is 0.49, $p > 0.62$ (0.62, $p > 0.53$). In Israel, this might be explained by the setting: the participants took the experiment from their homes, online. Consequently, they might have responded differently than they would in the lab. It is also possible that students who were enrolled in the university/college during the COVID-19 pandemic were less exposed to economic ideas because they studied from home. It is not clear that they invested the same level of effort as students who attended regular in-class lectures. Another possibility is that the high level of unemployment during the COVID-19 lockdowns might have affected the participants' perceptions of unemployment.

More relevant for our hypotheses, however, we find that after being primed with communal values, both economics and non-economics students retained more workers than participants in the control treatment. This result holds for both Israel and the U.S.

To test this more formally, we estimate a series of OLS regressions with robust standard errors, similar to the regressions we report in Table 2. In all regressions, the dependent variable is the number of workers retained. In the baseline model, the only independent variables are dummies for the communal treatment, and for economics students. For the Israeli data, we cluster the standard errors according to the institution in which the participants studied. For the U.S. data, we use robust standard errors.

Column 1 (4) of Table 3 reports the results for the Israeli (U.S.) data. We find that in Israel (the U.S.), participants in the communal treatment retained 11.84, $p < 0.01$ (8.37, $p < 0.02$) more workers than participants in the control treatment. Thus, exposure to cues with communal values leads to a significant increase in the number of workers retained.

In column 2, we add further controls: woman, married, religious, employment, voting left-wing/democrats, parents with academic degrees, age, and the number of words recalled. All variables are defined as above. In Israel, the coefficients of all the added independent variables are not statistically significant. In the U.S., the coefficient of the participants' age, 0.87 ($p < 0.01$), is positive and statistically significant. The main result, however, remains unchanged: participants who were primed with communal values retained more workers than participants in the control treatment. In Israel, they retained 12.44 ($p < 0.01$) more workers. In the U.S., they retained 8.56 ($p < 0.02$) more workers.

In column 3, we add an interaction term between priming by communal values and being economics students. In Israel, the coefficient of the main effect of communal values remains positive, 8.16, and statistically significant ($p < 0.01$). In the U.S., the size of the coefficient of the main effect of communal values remains almost unaffected, 8.66, and marginally significant ($p < 0.07$). In Israel, the coefficient of the interaction term is also positive, 7.24, but it is not statistically significant ($p > 0.10$). In the U.S., the coefficient of the interaction term, $-0.23$, is small, negative, and not statistically significant ($p > 0.97$). It, therefore, seems that priming by communal values has a positive effect on the number of workers retained, and this effect is not different between economics students and other students.

## 7. Robustness

We run several robustness tests. First, we added to the regressions in Tables 2 and 3 fixed effects for the fields of study, and for the institution at which the experiment was conducted to control for heterogeneity between students that study fields other than economics, and for possible heterogeneity between students from different institutions.

Second, in experiment 1 we have data on students in their first week of studies at the university. This allows us to test for the indoctrination effect vs. the selection effect: students in their first week of studies were not indoctrinated yet by their professors. Therefore, if the differences between economics students and other students are due to indoctrination, then the differences that we find should be more pronounced among experienced students than among freshmen in their first week of school. However, we find evidence only in favor of selection: economics students in their first week of studies lay off as many workers as more experienced economics students. Most importantly, the main result remains unaffected; adding these controls does not have a significant effect on the results. For more details about these robustness tests, see the appendix.

## 8. Conclusions

Previous research suggests that in experimental settings, economics students tend to maximize profits in situations in which other participants adhere to social norms such as fairness and equality (Frank et al., 1993; Frank et al., 1996; Selten and Ockenfels, 1998; Cadsby and Maynes, 1998; Frank and Schulze, 2000). It has been also reported that years after graduation, economists still have different beliefs and attitudes than non-economists (Caplan, 2002; van Dalen, 2019). These findings are usually explained by stable personality traits, which are the outcome of either natural inclination, i.e., selection, or economic training, i.e., indoctrination (Frey et al., 1993).

Some studies, however, find no difference between economists and non-economists (Yezer et al., 1996; Laband and Beil, 1999). We offer a new explanation for these contradictory findings: we argue that the behavior of both economists and non-economists could be affected by their interpretation of the context.[13]

We test this hypothesis by conducting two experiments in which we manipulate the cues that the participants receive before their decision-making. We find that when participants receive cues signaling that the decision has an economic context, both economics and non-economics students tend to maximize profits. When the participants receive cues emphasizing social norms, on the other hand, both economics and non-economics students are less likely to maximize profits.

---

[13] Research in psychology suggests that there is a natural tendency to explain behavior by focusing on personality traits rather than on situational factors (Ross, 1977). See also Rubin (2003).

Thus, while the literature explains the behavior of economics students by focusing on their personality traits, our results suggest that the explanation may also have to do with contextual cues. In experiments in which both economics and non-economics students participate, one must consider that economics students are a priori more likely than other students to pay attention to cues related to markets and market norms because that is how they are trained. Consequently, they are more likely to interpret the experiment as set in a business environment and respond accordingly. Therefore, even if there are no ex-ante differences between the preferences of economics and non-economics students, economics students are likely to make different choices than non-economics students.

Before concluding, we note two issues that deserve further study. First, consistent with the previous research, in our first experiment, we find a baseline difference between economics and non-economics students. The analysis we run (see the appendix), suggests that this difference exists between economics and non-economics students in their first week of studies. In other words, this difference is likely to be an outcome of selection rather than indoctrination (Frey, 1993).

However, we do not find this difference in our second experiment. We do not have a satisfactory explanation for this finding, but we suspect that this may have to do with the aftermath of COVID-19. The outbreak of the pandemic led to high unemployment rates and a debate about economic policies. This might have changed the cues that economics students receive. Further research is needed to explore this issue in depth.

Second, in our first experiment, after being primed with economic terms, non-economics students chose to maximize profits to about the same extent as economics students who were not primed with economic terms. It would be interesting to see the outcome if we were to use stronger priming cues. Future research could consider the effects of using stronger cues or using cues only for economics students, or only for non-economics students. This could help us determine whether by manipulating the cues that each group receives, we can make non-economics students act like profit maximizers to a greater extent than economics students, and vice-versa.

Table 1. Summary statistics

| | A. Experiment 1 | | B. Experiment 2 | |
|---|---|---|---|---|
| | Israel | U.S. | Israel | U.S. |
| Workers retained | 133.8 | 153.7 | 144.3 | 152.6 |
| % Economics students | 65.6% | 40.4% | 59.0% | 44.8% |
| % Women | 59.9% | 37.4% | 53.0% | 46.2% |
| % Married | 10.3% | 7.1% | 36.6% | 6.1% |
| % Religious | 46.7% | | 22.3% | |
| % Political left/ % Voting Democrats | 10.8% | 40.4% | 13.3% | 30.2% |
| % Employed | 58.3% | 79.8% | 80.9% | 73.1% |
| % First week of studies | 52.8% | 0.0% | 0.0% | 0.0% |
| % having parents with academic degrees | 45.0% | 26.3% | 22.3% | 33.0% |
| Age | 23.3 | 22.3 | 30.1 | 21.9 |
| Words found in puzzle | 8.9 | 10.0 | 8.7 | 12.5 |
| Observations | 544 | 99 | 383 | 212 |

Notes:
The table presents summary statistics of the participants in the two experiments. Panel A gives the summary statistics of the participants in experiment 1. Panel B gives the summary statistics of the participants in experiment 2. Workers retained is the average response to the question about how many workers the participants chose to retain. % Economics students is the % of students studying economics, accounting, business administration, banking and finance, or management. % women is the % of women. % married is the % of married participants. % religious is the % of participants that identify themselves as either religious or ultra-religious. % political left/% voting democrats is the % of participants that vote for center-left/left-wing parties (Israeli data), or that vote for the Democratic party (U.S. data). % employed is the % of participants that work either part or full-time. Age is the average age of the participants. % first week of studies is the % of participants that took part in the experiment while in their first week of studies. % having parents with academic degrees is the % of the participants that both their parents have academic degrees. Age is the participants' age. Words found in the puzzle are the average number of words that the participants found in the puzzles.

Table 2. The number of workers retained in experiment 1

| | Israel | | | U.S | | |
|---|---|---|---|---|---|---|
| | (1) | (2) | (3) | (4) | (5) | (6) |
| Economic treatment | −10.69*** | −10.22*** | −12.47*** | −10.27** | −10.84** | −10.37* |
| | (2.165) | (2.173) | (2.025) | (4.813) | (4.759) | (6.232) |
| Economics student | −13.18*** | −12.871 | −14.56*** | −7.09 | −6.42 | −5.81 |
| | (3.277) | (3.588) | (3.908) | (4.893) | (5.005) | (5.830) |
| Woman | | −4.32 | −4.21 | | −5.72 | −5.72 |
| | | (3.578) | (3.570) | | (5.400) | (5.443) |
| Married | | 0.59 | 0.60 | | −2.78 | −2.68 |
| | | (5.238) | (5.220) | | (9.029) | (9.008) |
| Religious | | 0.05 | −0.06 | | | |
| | | (3.575) | (3.604) | | | |
| Voting left-wing/democrats | | 7.11* | 7.36* | | 2.03 | −2.011 |
| | | (3.652) | (3.677) | | (4.564) | (4.586) |
| Employment | | −4.871** | −4.78** | | 12.52* | 12.51* |
| | | (2.242) | (2.278) | | (6.591) | (6.643) |
| Parents with academic degrees | | 0.50 | 0.56 | | −3.89 | −4.01 |
| | | (3.274) | (3.273) | | (5.520) | (5.464) |
| Age | | 0.49 | 0.49 | | −0.58 | −0.58 |
| | | (0.360) | (0.359) | | (0.589) | (0.603) |
| # of words found in puzzle | | −0.30 | −0.30 | | −0.08 | −0.07 |
| | | (0.357) | (0.357) | | (0.876) | (0.887) |
| Economic treatment × Economics student | | | 3.38 | | | −1.21 |
| | | | (3.288) | | | (10.091) |
| Constant | 148.00*** | 143.24*** | 144.23*** | 161.58*** | 169.56*** | 169.25*** |
| | (3.213) | (11.783) | (11.778) | (3.135) | (17.622) | (18.366) |
| $R^2$ | 0.062 | 0.086 | 0.087 | 0.071 | 0.143 | 0.162 |
| Observations | 544 | 538 | 538 | 99 | 99 | 99 |

Notes:
The table presents the results of OLS regressions with standard errors clustered at the sessions' level. The dependent variable is the number of workers retained. Economic treatment is a dummy variable that equals 1 if the participant took part in the economic treatment and 0 if s/he participated in the control treatment. Economics students is a dummy that equals 1 if the participant is an economics student, and 0 otherwise. Woman is a dummy that equals 1 if the participant is a woman and 0 otherwise. Married is a dummy that equals 1 if the participant is married, and 0 otherwise. Religious is a dummy that equals 1 if the participant identifies himself as religious or ultra-religious, and 0 otherwise. Voting left-wing/democrats is a dummy that equals 1 if an Israeli (U.S.) participant responded that s/he votes for left or center-left parties (votes for the Democratic party), and 0 otherwise. Employment is a dummy that equals 1 if the participant works either part or full-time, and 0 otherwise. Parents with academic degrees is a dummy that equals 1 if the participant's both parents have academic degrees, and 0 otherwise. Age is the participant's age in years. # of words found in the puzzle is the number of words that the participants found in the puzzle. * $p < 10\%$, ** $p < 5\%$, *** $p < 1\%$.

Table 3. The number of workers retained in experiment 2

| | Israel | | | U.S | | |
|---|---|---|---|---|---|---|
| | (1) | (2) | (3) | (4) | (5) | (6) |
| Communal treatment | 11.84*** (1.679) | 12.44*** (1.739) | 8.16*** (2.079) | 8.37** (3.449) | 8.56** (3.468) | 8.66* (4.577) |
| Economics student | 0.92 (3.367) | 0.61 (2.8223) | −2.96 (2.762) | 2.57 (3.475) | 3.71 (3.553) | 3.82 (5.136) |
| Woman | | 2.28 (2.397) | 2.10 (2.520) | | 0.80 (3.716) | 0.80 (3.722) |
| Married | | −2.84 (5.784) | −3.10 (5.758) | | −1.89 (7.376) | −1.88 (7.427) |
| Religious | | −1.42 (3.996) | −1.22 (4.075) | | | |
| Voting left-wing/democrats | | −1.62 (5.724) | −1.33 (5.869) | | −4.62 (3.859) | −4.27 (3.859) |
| Employment | | 7.33 (6.575) | 7.46 (6.650) | | 0.01 (3.851) | 0.02 (3.916) |
| Parents with academic degrees | | −2.58 (4.262) | −2.91 (4.210) | | 4.52 (3.590) | 4.52 (3.59) |
| Age | | 0.27 (0.201) | 0.28 (0.201) | | 0.87*** (0.332) | 0.87** (0.334) |
| # of words found in puzzle | | 0.49 (0.510) | 0.49 (0.501) | | −0.656 (0.443) | −0.656 (0.444) |
| Economic treatment × Economics student | | | 7.24 (4.424) | | | −0.23 (7.034) |
| Constant | 137.868*** (3.920) | 120.61*** (8.768) | 122.74*** (9.090) | 147.06*** (3.000) | 135.10*** (10.320) | 135.06*** (10.479) |
| $R^2$ | 0.033 | 0.054 | 0.057 | 0.029 | 0.074 | 0.074 |
| Observations | 383 | 383 | 383 | 212 | 212 | 212 |

Notes:
The table presents the results of OLS regressions with standard errors clustered at the session level. The dependent variable is the number of workers retained. Communal treatment is a dummy variable that equals 1 if the participant took part in the communal treatment and 0 if s/he participated in the control treatment. Economics student is a dummy that equals 1 if the participant is an economics student, and 0 otherwise. Woman is a dummy that equals 1 if the participant is a woman and 0 otherwise. Married is a dummy that equals 1 if the participant is married, and 0 otherwise. Religious is a dummy that equals 1 if the participant identifies himself as religious or ultra-religious, and 0 otherwise. Voting left-wing/democrats is a dummy that equals 1 if an Israeli (U.S.) participant responded that s/he votes for left or center-left parties (votes for the Democratic party), and 0 otherwise. Employment is a dummy that equals 1 if the participant works part-time or full-time, and 0 otherwise. Parents with academic degrees is a dummy that equals 1 if a participant's both parents have academic degrees, and 0 otherwise. Age is the participant's age in years. # of words found in the puzzle is the number of words that the participants found in the puzzle. * $p < 10\%$, ** $p < 5\%$, *** $p < 1\%$.

Figure 1. The number of workers retained in experiment 1

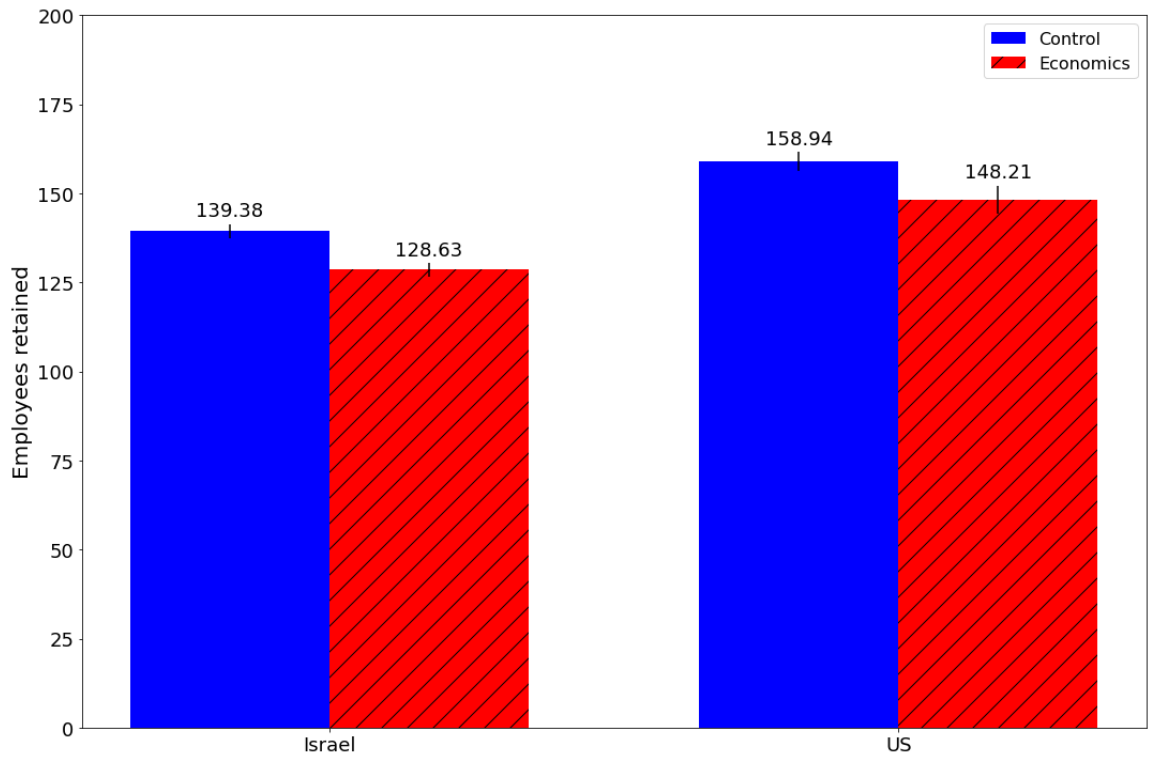

Notes:
The average number of workers retained by the participants. Vertical lines indicate the standard errors of the means.

Figure 2. The number of workers retained in experiment 1: Economics students
vs. other students

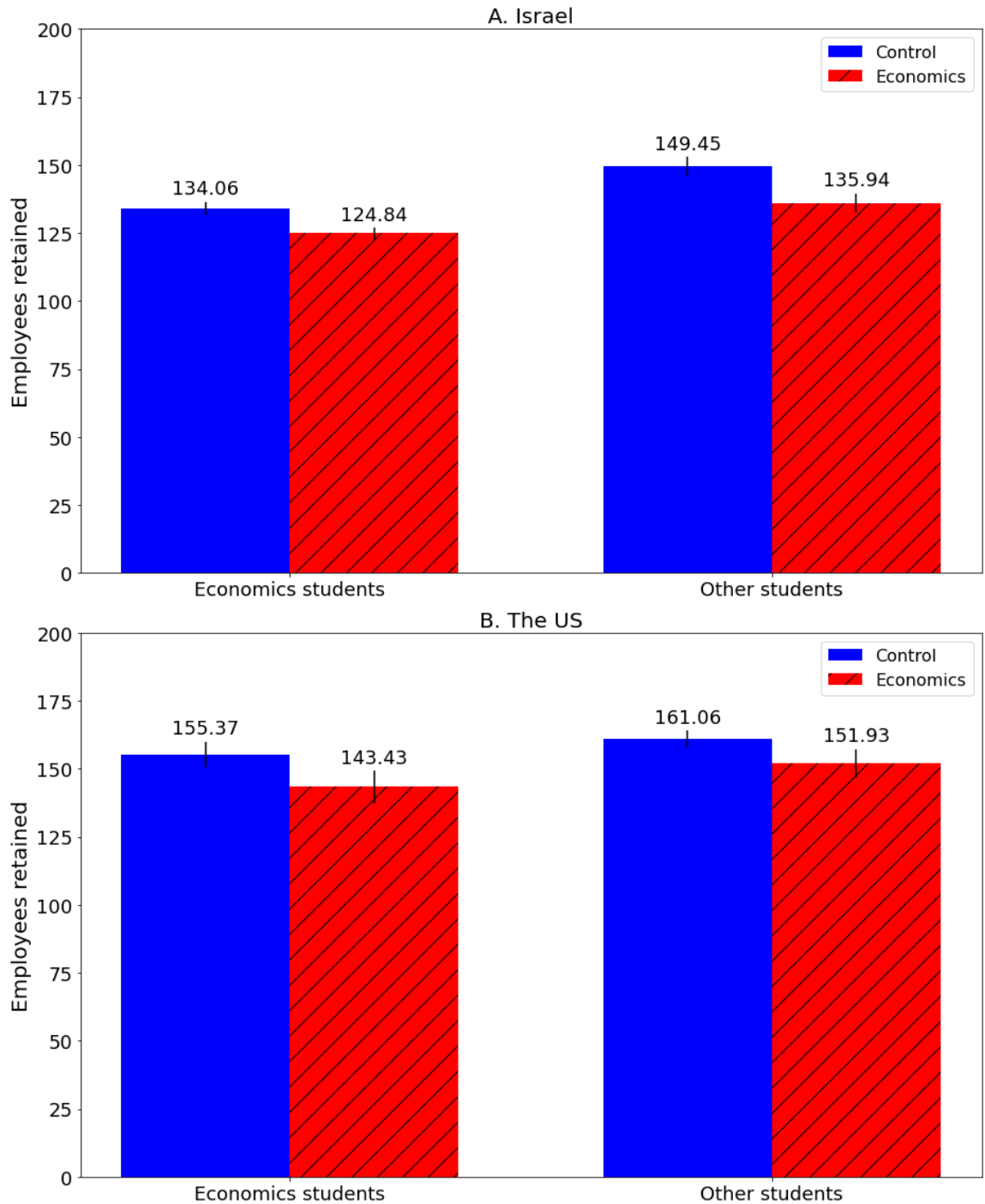

Notes:
The number of workers retained by the participants. Panel A gives the results for the Israeli data.
Panel B gives the results for the U.S. data. Vertical lines indicate the standard errors of the means.

Figure 3. The number of workers retained in experiment 2

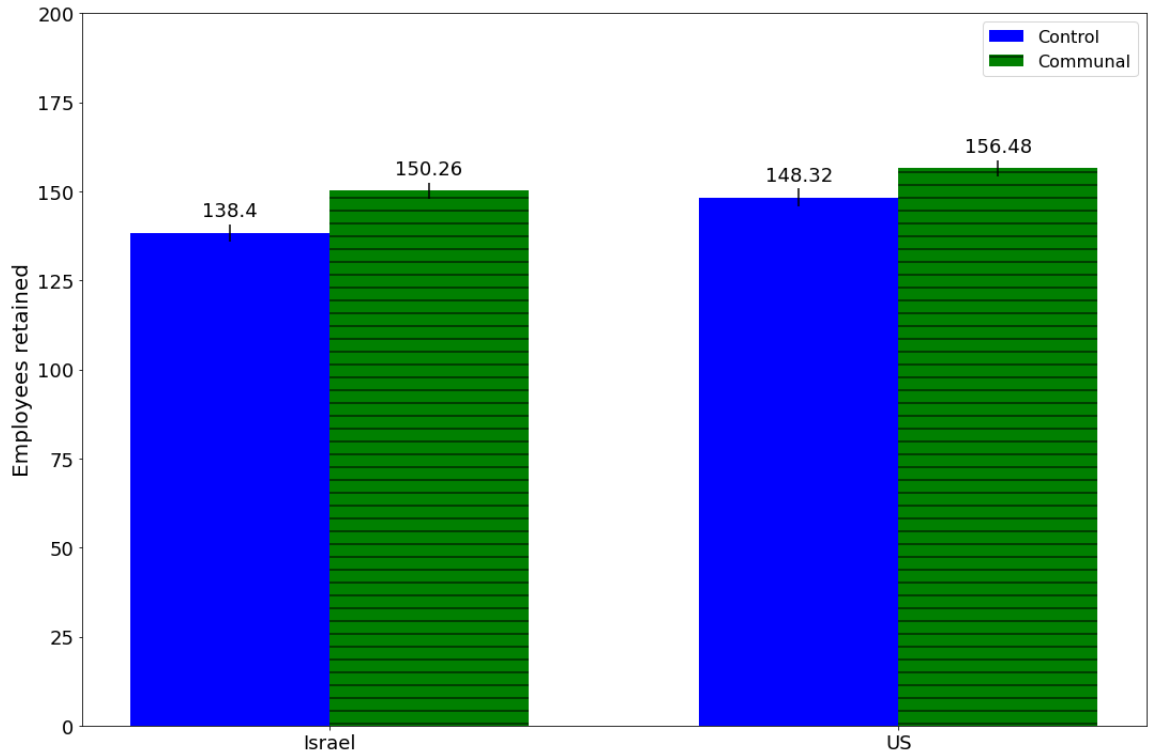

Notes:
The average number of workers retained by the participants. Vertical lines indicate the standard errors of the means.

Figure 4. The number of workers retained in experiment 2: Economics students vs. other students

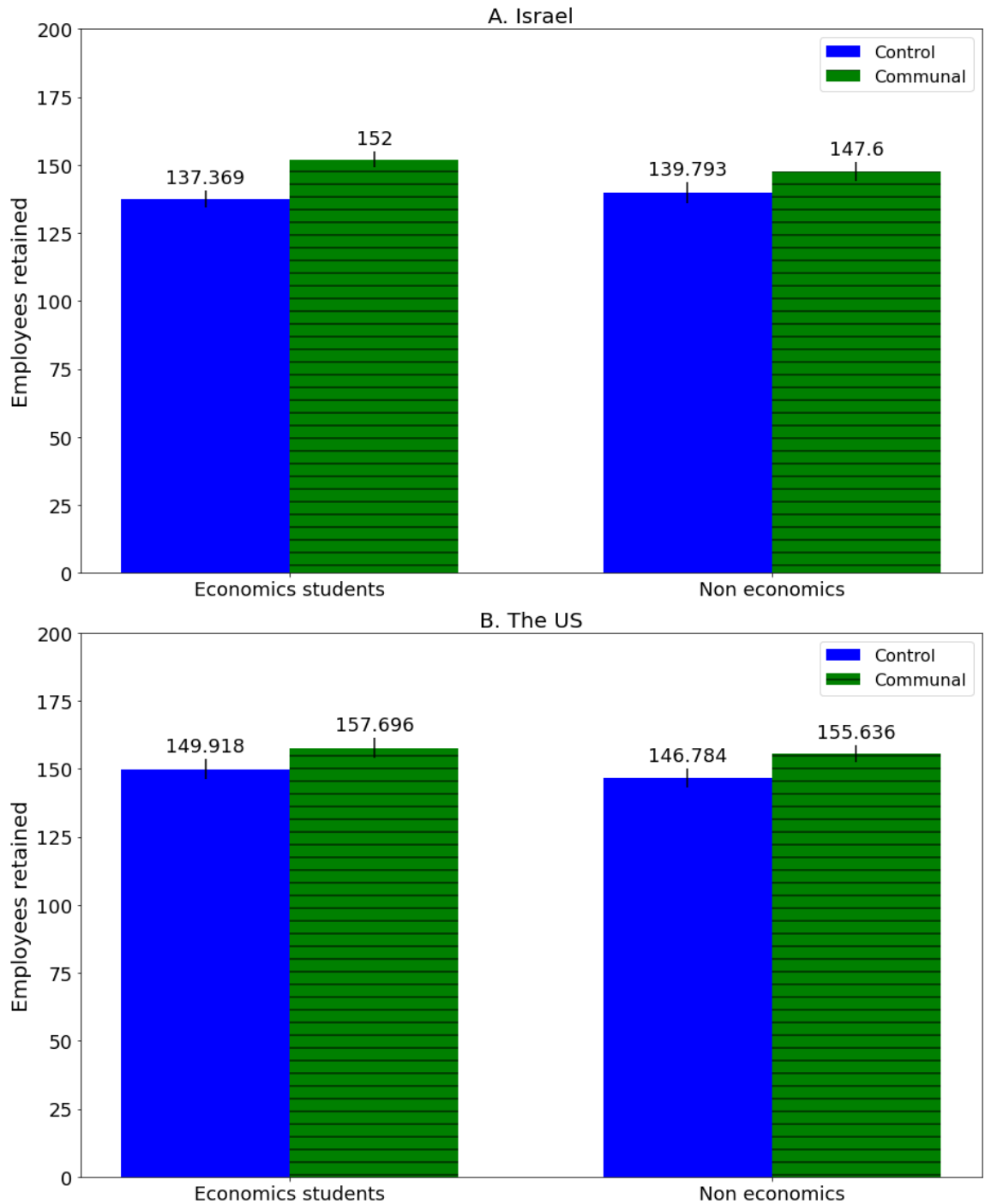

Notes:
The number of workers retained by the participants. Panel A gives the results for the Israeli data. Panel B gives the results for the U.S. data. Vertical lines indicate the standard errors of the means.

**Online Supplementary Web Appendix**

# Large Effects of Small Cues:
# Priming Selfish Economic Decisions*

**Avichai Snir\*\***
Department of Economics
Bar-Ilan University
avichai.snir@biu.ac.il

**Dudi Levy**
Department of Economics
Bar-Ilan University
dudilevi1985@gmail.com

**Dian Wang**
Alvarez College of Business
University of Texas at San Antonio
dian.wang@utsa.edu

**Haipeng (Allan) Chen**
Tippie College of Business
University of Iowa
haipeng-chen@uiowa.edu

**Daniel Levy**
Department of Economics
Bar-Ilan University,
Department of Economics
Emory University,
ICEA, ISET at TSU, and RCEA
Daniel.Levy@biu.ac.il

April 21, 2024

# Table of Contents

# Appendix A. Experiment procedures

## Experiment 1. Control treatment (Same as in experiment 2)

Word search task - 5 min to find 15 terms

| | | | |
|---|---|---|---|
| Laptop | _______ | Street | _______ |
| Toolbox | _______ | Glasses | _______ |
| Garbage bin | _______ | Electric gate | _______ |
| Book | _______ | Desk | _______ |
| Game console | _______ | Carpet | _______ |
| Window | _______ | Night lamp | _______ |
| Umbrella | _______ | Car | _______ |
| Toothbrush | _______ | | |

| E | B | T | B | O | O | K | M | L | A | I | O | L | I | T | S | U |
| R | U | M | B | R | E | L | L | A | T | E | A | O | L | R | T | T |
| T | G | D | F | G | V | S | V | A | C | P | X | K | I | E | R | O |
| G | C | A | R | G | C | O | D | D | T | C | V | R | T | T | E | O |
| I | R | G | M | E | K | R | W | O | T | A | C | E | R | A | E | T |
| N | E | F | E | E | A | E | P | O | A | P | E | W | A | G | T | H |
| I | M | V | V | A | C | C | U | K | C | M | W | D | G | C | G | B |
| G | O | L | I | N | R | O | M | O | G | A | A | F | L | I | F | R |
| H | F | F | R | A | N | D | N | R | O | I | I | G | A | R | I | U |
| T | W | T | E | P | R | A | C | S | D | L | U | C | S | T | U | S |
| L | M | O | X | O | N | I | E | R | O | K | D | X | S | C | H | H |
| A | B | N | D | R | T | G | A | S | G | L | X | Z | E | E | V | W |
| M | M | A | L | N | P | A | C | K | A | F | E | N | S | L | E | G |
| P | U | R | U | K | I | R | L | I | A | T | W | J | A | E | R | J |
| R | U | L | L | E | M | W | T | O | L | O | B | M | B | B | T | L |
| B | L | O | O | F | E | A | D | G | X | O | B | L | O | O | T | P |
| W | D | E | S | K | R | G | A | R | B | A | G | E | B | I | N | O |

**I found __________ of the 15 words listed above.**

**Questions**

**(There is no right or wrong answer.)**

**Question 1**

Assume that you are the vice president of ILJK Company. The company provides extermination services and employs administrative workers who cannot be fired and 196 non-permanent workers who do the actual extermination work and can be fired. The company was founded 5 years ago and is owned by three families. The work requires only a low level of skills, so each worker requires only one week of training. All of the company's employees have been with the company for three to five years. The company pays its workers more than the minimum wage. A worker's wage, which includes overtime, amounts to between $1,800 to $2,000 per month. The company provides its employees with all the benefits required by law.

Until recently, the company was very profitable. As a result of the continuing recession, however, there has been a significant drop in profits though the company is still in the black. You will soon be attending a meeting of the management at which a decision will be made as to how many workers to lay off. ILJK's Finance Department has prepared the following forecast of annual profits:

| Number of workers who will continue to be employed | Expected annual profit in $ Millions |
| --- | --- |
| 100 (96 workers will be laid off) | Profit of 2 |
| 144 (52 workers will be laid off) | Profit of 1.6 |
| 170 (26 workers will be laid off) | Profit of 1 |
| 196 (no layoffs) | Profit of 0.4 |

I will recommend continuing to employ ☐ 100   ☐ 144   ☐ 170   ☐ 196   workers in the company

**Question 2**

What do you think would be the choice of a real vice president in Question 1? I think that he would recommend continuing to employ

☐ 100   ☐ 144   ☐ 170   ☐ 196   workers in the company

**Question 3**

A hardware store has been selling snow shovels for $17.99. The morning after a large snowstorm, the store raises the price to $22.99. This action is:

☐ Completely fair   ☐ Acceptable   ☐ Unfair   ☐ Very unfair

**Question 4**

At a sight-seeing point, reachable only on foot, a well has been tapped. The bottled water is sold to thirsty hikers. The price is $1.49 per bottle. Daily production and therefore the stock are 100 bottles. On a particularly hot day, the supplier raises the price to $2.99 per bottle. This action is:

☐ Completely fair   ☐ Acceptable   ☐ Unfair   ☐ Very unfair

**Question 5**

The gap between the rich and the poor should be reduced significantly:

☐ Completely agree   ☐ Somewhat agree   ☐ Neutral   ☐ Somewhat disagree   ☐ Completely disagree

**Question 6**

Environment-friendly corporations should be rewarded by the government:
☐ Completely agree   ☐ Somewhat agree   ☐ Neutral   ☐ Somewhat disagree   ☐ Completely disagree

**Age:** ______

**Gender:**
☐ Male   ☐ Female

**Marital status:**
☐ Single   ☐ Married   ☐ Divorced   ☐ Widower   ☐ Other:________

**Racial or ethnic origin:**
☐ American Indian or other Native American   ☐ Asian or Pacific Islander

☐ Black or African American   ☐ Caucasian (other than Hispanic)   ☐ Hispanic   ☐ Other

**Status in the college:**
☐ Freshman   ☐ Sophomore   ☐ Junior   ☐ Senior   ☐ Graduate student   ☐ Other

**Your major (or anticipated major):** _______________________________

**Did either of your parents graduate from college?**
☐ No   ☐ Yes, both parents   ☐ Yes, mother only   ☐ Yes, father only

**Do you work?**
☐ No   ☐ Yes, part-time   ☐ Yes, full-time

**Have you taken any courses in economics?**
☐ No   ☐ Yes, 1–2   ☐ Yes, 3–4   ☐ Yes, more than 4

**Do you describe yourself as:**
☐ Democrat   ☐ Republican   ☐ Independent   ☐ Other / I don't know

**Hanging out with friends (hours per week):**
☐ 1   ☐ 2   ☐ 3   ☐ 4   ☐ 5   ☐ 6 or more

**Do you recycle any one of the following: plastic/paper/newspaper/glass/batteries/etc.**

☐ Yes  ☐ No

**Did you volunteer in any setting during the last 12 months?**

☐ Yes  ☐ No

# Experiment 1. Economics treatment

**Word search mission- 5 min to find 15 terms**

| | | | |
|---|---|---|---|
| Inflation | ________ | Prime Rate | ________ |
| Recession | ________ | Income Tax | ________ |
| Price | ________ | Cost | ________ |
| Exchange Rate | ________ | Demand | ________ |
| Budget Deficit | ________ | Minimum Wage | ________ |
| Monopoly | ________ | Supply | ________ |
| Premia | ________ | Market | ________ |
| Unemployment | ________ | | |

| Q | Y | K | D | E | M | A | N | D | T | U | L | I | B | O | C | E |
|---|---|---|---|---|---|---|---|---|---|---|---|---|---|---|---|---|
| I | U | N | E | M | P | L | O | Y | M | E | N | T | S | X | O | R |
| R | T | B | R | A | N | D | I | O | T | F | S | Y | E | K | M | E |
| B | L | L | Q | U | A | N | N | E | L | H | O | T | X | E | P | M |
| U | A | U | A | S | L | O | I | A | P | L | I | O | C | G | E | A |
| D | N | V | B | S | P | P | T | H | P | O | Y | L | H | A | A | S |
| G | D | Q | N | O | A | I | L | J | U | A | T | O | A | W | I | U |
| E | B | Z | L | S | O | D | I | R | T | N | W | Y | N | M | B | P |
| T | R | Y | A | N | Q | T | M | W | F | D | C | I | G | U | I | P |
| D | U | G | I | L | S | W | Q | A | E | G | A | T | E | M | V | L |
| E | X | E | T | B | I | E | C | I | R | P | R | R | I | P | Y |
| F | B | A | R | E | T | R | I | I | E | K | P | A | A | N | R | L |
| I | R | E | C | E | S | S | I | O | N | A | E | C | T | I | E | U |
| C | P | R | D | I | U | C | E | U | M | A | N | T | E | M | M | S |
| I | I | B | E | T | A | R | E | M | I | R | P | E | Y | R | I | B |
| T | O | X | A | T | E | M | O | C | N | I | P | M | O | N | A | U |
| X | C | A | T | M | A | R | D | U | N | A | L | C | O | S | T | G |

**I found ___________ of the 15 words listed above.**

**Questions**

**(There is no right or wrong answer.)**

**Question 1**

Assume that you are the vice president of ILJK Company. The company provides extermination services and employs administrative workers who cannot be fired and 196 non-permanent workers who do the actual extermination work and can be fired. The company was founded 5 years ago and is owned by three families. The work requires only a low level of skills, so each worker requires only one week of training. All of the company's employees have been with the company for three to five years. The company pays its workers more than the minimum wage. A worker's wage, which includes overtime, amounts to between $1,200–$1,440, per month. The company provides its employees with all the benefits required by law.

Until recently, the company was very profitable. As a result of the continuing recession, however, there has been a significant drop in profits though the company is still in the black. You will soon be attending a meeting of the management at which a decision will be made as to how many workers to lay off. ILJK's Finance Department has prepared the following forecast of annual profits:

| Number of workers who will continue to be employed | Expected annual profit in $ Millions |
|---|---|
| 100 (96 workers will be laid off) | Profit of 2 |
| 144 (52 workers will be laid off) | Profit of 1.6 |
| 170 (26 workers will be laid off) | Profit of 1 |
| 196 (no layoffs) | Profit of 0.4 |

I will recommend continuing to employ ☐ 100  ☐ 144  ☐ 170  ☐ 196  workers in the company

**Question 2**

What do you think would be the choice of a real vice president in Question 1? I think that he would recommend continuing to employ

☐ 100  ☐ 144  ☐ 170  ☐ 196  workers in the company

**Question 3**

A hardware store has been selling snow shovels for $17.99. The morning after a large snowstorm, the store raises the price to $22.99. This action is:

☐ Completely fair   ☐ Acceptable   ☐ Unfair   ☐ Very unfair

**Question 4**

At a sight-seeing point, reachable only on foot, a well has been tapped. The bottled water is sold to thirsty hikers. The price is $1.49 per bottle. Daily production and therefore the stock are 100 bottles. On a particularly hot day, the supplier raises the price to $2.99 per bottle. This action is:

☐ Completely fair   ☐ Acceptable   ☐ Unfair   ☐ Very unfair

**Question 5**

The gap between the rich and the poor should be reduced significantly:

☐ Completely agree   ☐ Somewhat agree   ☐ Neutral   ☐ Somewhat disagree   ☐ Completely disagree

**Question 6**

Environment-friendly corporations should be rewarded by the government:

☐ Completely agree  ☐ Somewhat agree  ☐ Neutral  ☐ Somewhat disagree  ☐ Completely disagree

**Age:** ______

**Gender:**

☐ Male  ☐ Female

**Marital status:**

☐ Single  ☐ Married  ☐ Divorced  ☐ Widower  ☐ Other:________

**Racial or ethnic origin:**

☐ American Indian or other Native American  ☐ Asian or Pacific Islander

☐ Black or African American  ☐ Caucasian (other than Hispanic)  ☐ Hispanic  ☐ Other

**Status in the college:**

☐ Freshman  ☐ Sophomore  ☐ Junior  ☐ Senior  ☐ Graduate student  ☐ Other

**Your major (or anticipated major):** ___________________________________

**Did either of your parents graduate from college?**

☐ No  ☐ Yes, both parents  ☐ Yes, mother only  ☐ Yes, father only

**Do you work?**

☐ No  ☐ Yes, part-time  ☐ Yes, full-time

**Have you taken any courses in economics?**

☐ No  ☐ Yes, 1–2  ☐ Yes, 3–4  ☐ Yes, more than 4

**Do you describe yourself as:**

☐ Democrat  ☐ Republican  ☐ Independent  ☐ Other / I don't know

**Hanging out with friends (hours per week):**

☐ 1  ☐ 2  ☐ 3  ☐ 4  ☐ 5  ☐ 6 or more

**Do you recycle any one of the following: plastic/paper/newspaper/glass/batteries/etc.**

☐ Yes ☐ No

**Did you volunteer in any setting during the last 12 months?**

☐ Yes ☐ No

Experiment 2. Communal treatment

**Word search task - 5 min to find 15 terms**

| | | | |
|---|---|---|---|
| Equality | _______ | Kindhearted | _______ |
| Mutual trust | _______ | Social Norm | _______ |
| Kindness | _______ | Caring | _______ |
| Solidarity | _______ | Compassion | _______ |
| Altruism | _______ | Painful | _______ |
| Charity | _______ | Humane | _______ |
| Basic Needs | _______ | Help | _______ |
| Fairness | _______ | | |

| Q | Y | K | H | U | M | A | N | E | T | U | L | X | B | T | C | E |
|---|---|---|---|---|---|---|---|---|---|---|---|---|---|---|---|---|
| C | O | M | P | A | S | S | I | O | N | C | E | B | S | N | O | R |
| K | T | B | R | A | N | D | I | F | T | Q | S | Y | S | T | X | E |
| I | L | L | Q | U | A | N | C | E | U | H | O | T | O | E | P | M |
| N | A | U | A | S | L | H | I | A | P | L | I | O | C | T | E | A |
| D | N | V | B | S | A | P | L | H | P | O | Y | L | I | S | A | F |
| H | D | Q | N | R | A | I | L | J | U | A | T | O | A | U | I | A |
| E | B | Z | I | S | T | D | I | R | T | N | W | Y | L | R | B | I |
| A | R | T | A | Y | Q | T | F | W | F | D | C | I | N | T | I | R |
| R | Y | G | I | L | S | W | Q | P | E | G | A | T | O | L | V | N |
| T | S | S | E | N | D | N | I | K | A | K | R | R | A | C | E |
| E | B | A | R | E | T | R | I | I | E | I | P | A | M | U | C | S |
| D | A | L | T | R | U | I | S | M | N | A | N | C | F | T | A | S |
| T | P | R | D | I | U | C | E | U | M | A | N | F | D | U | R | I |
| P | 1 | B | A | S | I | C | N | E | E | D | S | F | U | M | I | B |
| A | O | N | Y | T | I | R | A | D | I | L | O | S | O | L | N | U |
| X | C | A | T | M | A | R | D | U | H | E | L | P | R | S | G | G |

**I found __________ of the 15 words listed above.**

**Questions**

**(There is no right or wrong answer.)**

**Question 1**

Assume that you are the vice president of ILJK Company. The company provides extermination services and employs administrative workers who cannot be fired and 196 non-permanent workers who do the actual extermination work and can be fired. The company was founded 5 years ago and is owned by three families. The work requires only a low level of skills, so each worker requires only one week of training. All of the company's employees have been with the company for three to five years. The company pays its workers more than the minimum wage. A worker's wage, which includes overtime, amounts to between $1,200–$1,440, per month. The company provides its employees with all the benefits required by law.

Until recently, the company was very profitable. As a result of the continuing recession, however, there has been a significant drop in profits though the company is still in the black. You will soon be attending a meeting of the management at which a decision will be made as to how many workers to lay off. ILJK's Finance Department has prepared the following forecast of annual profits:

| Number of workers who will continue to be employed | Expected annual profit in $ Millions |
|---|---|
| 100 (96 workers will be laid off) | Profit of 2 |
| 144 (52 workers will be laid off) | Profit of 1.6 |
| 170 (26 workers will be laid off) | Profit of 1 |
| 196 (no layoffs) | Profit of 0.4 |

I will recommend continuing to employ ☐ 100  ☐ 144  ☐ 170  ☐ 196  workers in the company

**Question 2**

What do you think would be the choice of a real vice president in Question 1? I think that he would recommend continuing to employ

☐ 100  ☐ 144  ☐ 170  ☐ 196  workers in the company

**Question 3**

A hardware store has been selling snow shovels for $17.99. The morning after a large snowstorm, the store raises the price to $22.99. This action is:

☐ Completely fair  ☐ Acceptable  ☐ Unfair  ☐ Very unfair

**Question 4**

At a sight-seeing point, reachable only on foot, a well has been tapped. The bottled water is sold to thirsty hikers. The price is $1.49 per bottle. Daily production and therefore the stock are 100 bottles. On a particularly hot day, the supplier raises the price to $2.99 per bottle. This action is:

☐ Completely fair  ☐ Acceptable  ☐ Unfair  ☐ Very unfair

**Question 5**

The gap between the rich and the poor should be reduced significantly:

☐ Completely agree  ☐ Somewhat agree  ☐ Neutral  ☐ Somewhat disagree  ☐ Completely disagree

**Question 6**

Environment-friendly corporations should be rewarded by the government:

☐ Completely agree  ☐ Somewhat agree  ☐ Neutral  ☐ Somewhat disagree  ☐ Completely disagree

**Age:** _______

**Gender:**

☐ Male  ☐ Female

**Marital status:**

☐ Single  ☐ Married  ☐ Divorced  ☐ Widower  ☐ Other:________

**Racial or ethnic origin:**

☐ American Indian or other Native American  ☐ Asian or Pacific Islander

☐ Black or African American  ☐ Caucasian (other than Hispanic)  ☐ Hispanic  ☐ Other

**Status in the college:**

☐ Freshman  ☐ Sophomore  ☐ Junior  ☐ Senior  ☐ Graduate student  ☐ Other

**Your major (or anticipated major):** _________________________________

**Did either of your parents graduate from college?**

☐ No  ☐ Yes, both parents  ☐ Yes, mother only  ☐ Yes, father only

**Do you work?**

☐ No  ☐ Yes, part-time  ☐ Yes, full-time

**Have you taken any courses in economics?**

☐ No  ☐ Yes, 1–2  ☐ Yes, 3–4  ☐ Yes, more than 4

**Do you describe yourself as:**

☐ Democrat  ☐ Republican  ☐ Independent  ☐ Other / I don't know

**Hanging out with friends (hours per week):**

☐ 1  ☐ 2  ☐ 3  ☐ 4  ☐ 5  ☐ 6 or more

**Do you recycle any one of the following: plastic/paper/newspaper/glass/batteries/etc.**

☐ Yes  ☐ No

**Did you volunteer in any setting during the last 12 months?**

☐ Yes  ☐ No

## Appendix B. Controlling for fields of study

In both, Israel and the US, we asked participants about their field of study, and we can therefore use their answers as controls in the regressions that we estimate.

In Table B1, we report the results of estimating regressions similar to the ones that we estimated in the paper. The dependent variable is the number of workers retained. The independent variables include: a dummy for the economic treatment, which equals 1 if the participant took part in the economics treatment and 0 otherwise, and a dummy for economics students, which equals 1 if the participant studied economics and 0 otherwise, Woman – a dummy that equals 1 if the participant is a woman, and 0 otherwise, married – a dummy that equals 1 if the participant is married, and 0 otherwise, religious – a dummy that equals 1 if the participant identified herself as religious or ultra-religious, and 0 otherwise, voting left-wing/democrats – a dummy that equals 1 if an Israeli (US) participant responded that s/he votes for left or center-left parties (votes for the Democratic party), and 0 otherwise, employment – a dummy that equals 1 if the participant works either part or full time, and 0 otherwise, academic – a dummy that equals 1 if both the participant's parents have academic degrees, and 0 otherwise, the participant's age in years, the number of words that the participants found in the puzzle and fixed effects for fields of study.

For the Israeli data, in experiment 1 we also add a fixed effect for the university at which the experiment took place (Bar-Ilan University vs. Tel-Aviv University). In experiment 2, we add fixed effects for the universities where the participants studied. We estimate the regression using OLS with robust standard errors, and we cluster the standard errors by sessions.

Column 1 reports the results of experiment 1 using Israeli data. We find that the values of the coefficient of interest remain similar to the values that we report in the paper. Participants in the economic treatment retained 10.36 ($p < 0.01$) fewer workers than participants in the control treatment. Column 3 reports the results for experiment 1 using the US data. The coefficient of the economic treatment is $-9.39$, and it is marginally significant ($p < 0.10$).

Column 3 reports the results of experiment 2 using Israeli data. Participants in the social treatment retained 11.64 ($p < 0.01$) more workers than participants in the control treatment. Column 4 reports the results of experiment 2 using the US data. The coefficient of the economic treatment is 7.81, and it is statistically significant ($p < 0.05$).

Table A1. Number of workers retained, controlling for the field of study and institution

| | Israel | | US | |
|---|---|---|---|---|
| | (1) | (2) | (3) | (4) |
| Economic treatment | −10.36*** | 11.64*** | −9.39* | 7.81** |
| | (2.207) | (2.652) | (5.485) | (3.461) |
| Economics student | −14.42*** | 4.05 | 9.23 | −4.52 |
| | (1.450) | (3.990) | (7.428) | (5.849) |
| Woman | −5.56 | 2.35 | −9.84 | −0.86 |
| | (4.745) | (2.172) | (6.429) | (3.886) |
| Married | −0.45 | −8.18 | −5.96 | −3.13 |
| | (5.132) | (7.450) | (9.495) | (7.414) |
| Religious | −0.52 | −3.17 | | |
| | (3.488) | (5.505) | | |
| Voting left-wing/democrats | 7.48** | 3.53 | −2.64 | −3.23 |
| | (3.510) | (4.950) | (5.219) | (4.224) |
| Employment | −5.47 | 1.20 | 9.37 | −0.97 |
| | (2.311) | (9.072) | (7.126) | (4.059) |
| Parents with academic degrees | −0.23 | −0.76 | −9.34 | 4.70 |
| | (3.187) | (4.242) | (6.849) | (3.662) |
| Age | 0.42 | 0.31 | −0.27 | 0.86** |
| | (0.406) | (0.377) | (0.853) | (0.342) |
| # of words found in puzzle | −0.36 | 0.33 | −0.149 | −0.74 |
| | (0.359) | (0.456) | (1.034) | (0.463) |
| Constant | 146.70*** | 141.28*** | 153.19*** | 133.48*** |
| | (11.332) | (20.114) | (20.856) | (11.277) |
| $R^2$ | 0.113 | 0.128 | 0.353 | 0.091 |
| Observations | 538 | 301 | 99 | 212 |

Notes
The table presents the results of OLS regressions with standard errors clustered at the sessions' level. The dependent variable is the number of employees retained. Economic treatment is a dummy variable that equals 1 if the participant took part in the economic treatment and 0 if s/he participated in the control treatment. Economics students is a dummy that equals 1 if the participant is an economics student, and 0 otherwise. Woman is a dummy that equals 1 if the participant is a woman, and 0 otherwise. Married is a dummy that equals 1 if the participant is married, and 0 otherwise. Religious is a dummy that equals 1 if the participant identifies himself as religious or ultra-religious, and 0 otherwise. Voting left-wing/democrats is a dummy that equals 1 if an Israeli (US) participant responded that s/he votes for left or center-left parties (votes for the Democratic party), and 0 otherwise. Employment is a dummy that equals 1 if the participant works either part or full-time, and 0 otherwise. Parents with academic degrees is a dummy that equals 1 if both parents of the participant have academic degrees, and 0 otherwise. Age is the participant's age in years. # of words found in the puzzle is the number of words that the participants found in the puzzle. All the regressions include fixed effects for the participants' majors. The regressions for the Israeli data also include fixed effects for the student's college/university. Column 1 gives the results of a regression using data from the first experiment conducted in Israel. Column 2 gives the results of a regression using data from the second experiment conducted in Israel. Column 3 gives the results of a regression using data from the first experiment conducted in the US. Column 4 gives the results of a regression using data from the second experiment conducted in the US.

* $p < 10\%$, ** $p < 5\%$, *** $p < 1\%$.

**Appendix C. Selection vs. indoctrination**

In the first experiment we conducted in Israel, we collected data from students in the first week of studies, as well as from more experienced students. This allows us to discriminate between selection and indoctrination hypotheses. Under the selection hypothesis, economics students in the first week of studies should make decisions similar to more experienced students. Under the indoctrination hypothesis, economics students should make different choices than other students only after some exposure to economic ideas and content.

In Table C1, we report the results of an OLS regression with robust standard errors, similar to the one we estimate in the paper. The dependent variable is the number of workers retained.

The independent variables include: a dummy for the economic treatment, which equals 1 if the participant took part in the economics treatment and 0 otherwise, and a dummy for economics students, which equals 1 if the participant studied economics and 0 otherwise, Woman – a dummy that equals 1 if the participant is a woman, and 0 otherwise, married – a dummy that equals 1 if the participant is married, and 0 otherwise, religious – a dummy that equals 1 if the participant identified herself as religious or ultra-religious, and 0 otherwise, voting left-wing/democrats – a dummy that equals 1 if an Israeli (US) participant responded that s/he votes for left or center-left parties (votes for the Democratic party), and 0 otherwise, employment – a dummy that equals 1 if the participant works either part or full time, and 0 otherwise, academic – a dummy that equals 1 if both parents of the participant have academic degrees, and 0 otherwise, the participant's age in years, and the number of words that the participants found in the puzzle.

In column 1, we also add control for students in their first week of studies (whether they major in economics or in another field). We find that students in their first week of studies retain 5.56 employees more than more experienced students ($p < 0.07$). This does not affect our main findings: Participants in the economics treatment are expected to retain 10.11 ($p < 0.01$) fewer employees than participants in the control treatment. In addition, economics students are expected to retain 13.80 ($p < 0.01$) fewer employees than non-economics students.

In column 2, we add an interaction for economics students in their first week of studies. This does not change the main findings: Participants in the economics treatment are expected to retain 10.12 ($p < 0.01$) fewer employees than participants in the control treatment. Economics students are expected to retain 13.33 ($p < 0.05$) fewer employees than non-economics students. The coefficient of economics students in the first week of studies is small, $-0.98$, and not statistically significant ($p > 0.87$).

It, therefore, seems that economics students retain fewer employees than non-economics students, and this difference exists even among students that only began their studies. Our results, therefore, support the selection hypothesis rather than the indoctrination hypothesis. In addition, controlling for that does not change the conclusion that participants in the economics treatment retain fewer employees.

Table C1. Number of workers retained, controlling for students in their first week of studies

|  | (1) | (2) |
|---|---|---|
| Economic treatment | −10.11*** | −10.12*** |
|  | (2.261) | (2.259) |
| Economics student | −13.80*** | −13.33** |
|  | (3.223) | (6.138) |
| Woman | −4.37 | −4.39 |
|  | (3.679) | (3.714) |
| Married | 1.16 | 1.20 |
|  | (5.206) | (5.189) |
| Religious | 0.42 | 0.44 |
|  | (3.542) | (3.548) |
| Voting left-wing/democrats | 7.11* | 7.17* |
|  | (3.574) | (3.600) |
| Employment | −4.09* | −4.13* |
|  | (2.140) | (2.215) |
| Parents with academic degrees | 0.18 | 0.21 |
|  | (3.164) | (3.207) |
| Age | 0.60 | 0.59 |
|  | (0.400) | (0.385) |
| # of words found in puzzle | −0.21 | −0.21 |
|  | (0.335) | (0.335) |
| First week of studies | 5.56* | 6.19 |
|  | (2.930) | (5.378) |
| Economics student × First week of studies |  | −0.98 |
|  |  | (6.225) |
| Constant | 137.06*** | 136.98*** |
|  | (13.172) | (13.385) |
| $R^2$ | 0.092 | 0.092 |
| Observations | 538 | 538 |

Notes

The table presents the results of OLS regressions with standard errors clustered at the sessions' level. The dependent variable is the number of employees retained. Economic treatment is a dummy variable that equals 1 if the participant took part in the economic treatment and 0 if s/he participated in the control treatment. Economics students is a dummy that equals 1 if the participant is an economics student, and 0 otherwise. Woman is a dummy that equals 1 if the participant is a woman, and 0 otherwise. Married is a dummy that equals 1 if the participant is married, and 0 otherwise. Religious is a dummy that equals 1 if the participant identifies himself as religious or ultra-religious, and 0 otherwise. Voting left-wing/democrats is a dummy that equals 1 if an Israeli (US) participant responded that s/he votes for left or center-left parties (votes for the Democratic party), and 0 otherwise. Employment is a dummy that equals 1 if the participant works either part or full-time, and 0 otherwise. Parents with academic degrees is a dummy that equals 1 if both parents of the participant have academic degrees, and 0 otherwise. Age is the participant's age in years. # of words found in the puzzle is the number of words that the participants found in the puzzle. The first week of studies is a dummy for students in their first week of studies. The regression uses data from the first experiment conducted in Israel.

* $p < 10\%$, ** $p < 5\%$, *** $p < 1\%$.

## Appendix D. The distribution of the responses

In the main part of the experiment, participants were asked to respond to the question, how many workers they would like to layoff. Tables D1 and D2 report the distribution of the responses in the first and second experiments, respectively.

From Table D1, it can be seen that when participants are primed with economic terms, they tend to retain fewer workers. This is true for both economic and non-economic students, and it happens in both Israel and the US samples. In particular, the share of participants that choose to profit maximize by retaining only 100 workers, increases in all groups. Among Israeli non-economics students, the share increases from 25.27% to 44.79%. Among Israeli economics students, the share increases from 40.12% to 55.68%. Among US non-economics students, the share increases from 0.00% to 14.81%. Among US economic students, the share increases from 5.26% to 23.81%.

From Table D2, it can be observed that when participants are primed with terms imbued with social values, they tend to retain more workers. Again, this is true for both economics and non-economics students, and it happens in both Israel and the US samples. In particular, the share of participants that choose to profit maximize by retaining 100 workers decreases in all groups. Among Israeli non-economics students, the share decreases from 30.00% to 17.57%. Among Israeli economics students, the share decreases from 36.36% to 20.87%. Among US non-economics students, the share decreases from 15.69% to 7.69%. Among US economics students, the share decreases from 14.29% to 8.70%.

Table D1. Distribution of responses, experiment 1

| Workers retained | Israel | | | |
|---|---|---|---|---|
| | **Non-priming** | | **Priming** | |
| | Non-economics | Economics | Non-economics | Economics |
| **100** | 25.27% | 40.12% | 44.79% | 55.68% |
| **144** | 30.77% | 37.79% | 20.83% | 28.65% |
| **170** | 24.18% | 14.53% | 23.96% | 10.81% |
| **196** | 19.78% | 7.56% | 10.42% | 4.86% |
| | **US** | | | |
| Workers retained | **Non-priming** | | **Priming** | |
| | Non-economics | Economics | Non-economics | Economics |
| **100** | 0.00% | 5.26% | 14.81% | 23.81% |
| **144** | 46.88% | 47.37% | 40.74% | 38.10% |
| **170** | 40.63% | 42.11% | 33.33% | 38.10% |
| **196** | 12.50% | 5.26% | 11.11% | 0.00% |

Table D2. Distribution of responses, experiment 2

| Workers retained | Israel | | | |
|---|---|---|---|---|
| | **Non-priming** | | **Priming** | |
| | Non-economics | Economics | Non-economics | Economics |
| **100** | 30.00% | 36.36% | 17.57% | 20.87% |
| **144** | 37.50% | 32.73% | 41.89% | 32.17% |
| **170** | 22.50% | 22.73% | 32.43% | 27.83% |
| **196** | 10.00% | 8.18% | 8.11% | 19.13% |
| | **US** | | | |
| Workers retained | **Non-priming** | | **Priming** | |
| | Non-economics | Economics | Non-economics | Economics |
| **100** | 15.69% | 14.29% | 7.69% | 8.70% |
| **144** | 50.98% | 44.90% | 49.23% | 39.13% |
| **170** | 29.41% | 34.69% | 30.77% | 36.96% |
| **196** | 3.92% | 6.12% | 12.31% | 15.22% |

## Appendix E. Summary statistics of the participants, by treatment

Tables E1–E4 give the summary statistics, by treatment, of the participants in the two experiments. Tables E1 and E2 give the summary statistics for experiment 1. Table E1 gives the summary statistics for the Israeli participants, and Table E2 for the US participants. Tables E3 and E4 give the summary statistics for experiment 2. Table E3 gives the summary statistics for the Israeli participants, and Table E4 for the US participants.

The figures in the tables suggest that the assignment of participants into treatment and control groups was indeed random, as the differences between the two groups in all treatments are usually small and not statistically significant.

However, in both of the Israeli treatments, there is a difference in the number of words found between participants in the treatment groups and the control groups. These differences should work "against us" in finding a priming effect because in both experiments it is the participants in the treatment groups that found fewer words than participants in the control group. Our results can therefore be interpreted as a conservative estimate of the effect of priming on the participants' choices.

Table E1. Israel, experiment 1 – economics treatment

|  | Control | Economics | *z*-value |
|---|---|---|---|
| Workers retained | 139.4 | 128.6 | 3.75*** |
| % Economics students | 65.4% | 65.8% | −0.11 |
| % Women | 60.1% | 59.8% | 0.07 |
| % Married | 12.2% | 8.5% | 1.39 |
| % Religious | 50.6% | 43.1% | 1.75* |
| % Political left/ % Voting Democrats | 10.6% | 11.0% | −0.14 |
| % Employed | 53.2% | 63.0% | −2.30** |
| % First week of studies | 53.2% | 52.3% | 0.21 |
| % having parents with academic degrees | 42.6% | 47.0% | −1.03 |
| Age | 23.4 | 23.2 | −0.39 |
| Words found in puzzle | 9.3 | 8.5 | 4.22*** |
| Observations | 263 | 281 |  |

Notes
The table presents summary statistics of the participants in the two experiments. Column 1 gives the summary statistics of the participants in the control treatment. Column 2 gives the summary statistics of the participants in the economics treatment. Column 3 gives the results of Wilcoxon rank sum test comparing the distributions. Workers retained is the average response to the question about how many workers the participants chose to retain. % Economics students is the % of students studying economics, accounting, business administration, banking and finance, or management. % women is the % of women. % married is the % of married participants. % religious is the % of participants that identify themselves as either religious or ultra-religious. % political left/% voting democrats is the % of participants that vote for center-left/left-wing parties (Israeli data), or that vote for the Democratic party (U.S. data). % employed is the % of participants that work either part or full-time. Age is the average age of the participants. % first week of studies is the % of participants that took part in the experiment while in their first week of studies. % having parents with academic degrees is the % of the participants that both their parents have academic degrees. Age is the participants' average age. Words found in the puzzle are the average number of words that the participants found in the puzzles.

* $p < 0.10$, ** $p < 0.05$, *** $p < 0.10$.

Table E2. US, experiment 1 – economics treatment

|  | Control | Economics | $z$-value |
|---|---|---|---|
| Workers retained | 158.9 | 148.2 | 1.71* |
| % Economics students | 37.3% | 43.8% | −0.66 |
| % Women | 37.3% | 37.5% | −0.03 |
| % Married | 7.8% | 6.3% | 0.73 |
| % Political left/ % Voting Democrats | 45.1% | 35.4% | 0.98 |
| % Employed | 82.4% | 77.1% | 0.65 |
| % having parents with academic degrees | 37.3% | 14.6% | 2.5** |
| Age | 22.4 | 22.3 | −0.70 |
| Words found in puzzle | 10.0 | 9.9 | 0.02 |
| Observations | 51 | 48 |  |

Notes
The table presents summary statistics of the participants in the two experiments. Column 1 gives the summary statistics of the participants in the control treatment. Column 2 gives the summary statistics of the participants in the economics treatment. Column 3 gives the results of Wilcoxon rank sum test comparing the distributions. Workers retained is the average response to the question about how many workers the participants chose to retain. % Economics students is the % of students studying economics, accounting, business administration, banking and finance, or management. % women is the % of women. % married is the % of married participants. % political left/% voting democrats is the % of participants that vote for center-left/left-wing parties (Israeli data), or that vote for the Democratic party (U.S. data). % employed is the % of participants that work either part or full-time. Age is the average age of the participants. % having parents with academic degrees is the % of the participants that both their parents have academic degrees. Age is the participants' average. Words found in the puzzle are the average number of words that the participants found in the puzzles.

* $p < 0.10$, ** $p < 0.05$, *** $p < 0.10$.

Table E3. Israel, experiment 2 – communal treatment

|  | Control | Communal | $z$-value |
|---|---|---|---|
| Workers retained | 138.4 | 150.3 | −3.45*** |
| % Economics students | 57.5% | 60.5% | −0.60 |
| % Women | 48.7% | 57.4% | −1.70* |
| % Married | 38.9% | 34.2% | 0.94 |
| % Religious | 25.4% | 20.5% | 1.1 |
| % Political left/ % Voting Democrats | 11.9% | 14.7% | −0.8 |
| % Employed | 80.8% | 81.1% | −0.06 |
| % having parents with academic degrees | 33.2% | 35.3% | −0.43 |
| Age | 29.8 | 30.3 | −0.35 |
| Words found in puzzle | 9.7 | 7.6 | 5.43*** |
| Observations | 193 | 190 |  |

Notes

The table presents summary statistics of the participants in the two experiments. Column 1 gives the summary statistics of the participants in the control treatment. Column 2 gives the summary statistics of the participants in the communal treatment. Column 3 gives the results of Wilcoxon rank sum test comparing the distributions. Workers retained is the average response to the question about how many workers the participants chose to retain. % Economics students is the % of students studying economics, accounting, business administration, banking and finance, or management. % women is the % of women. % married is the % of married participants. % religious is the % of participants that identify themselves as either religious or ultra-religious. % political left/% voting democrats is the % of participants that vote for center-left/left-wing parties (Israeli data), or that vote for the Democratic party (U.S. data). % employed is the % of participants that work either part or full-time. Age is the average age of the participants. % having parents with academic degrees is the % of the participants that both their parents have academic degrees. Age is the participants' average. Words found in the puzzle are the average number of words that the participants found in the puzzles.

* $p < 0.10$, ** $p < 0.05$, *** $p < 0.10$.

Table E4. US, experiment 2 – communal treatment

|  | Control | Economics | $z$-value |
|---|---|---|---|
| Workers retained | 148.3 | 156.5 | −2.17** |
| % Economics students | 49.0% | 41.1% | −14.5*** |
| % Women | 41.0% | 50.9% | 1.16 |
| % Married | 7.0% | 5.3% | −1.44 |
| % Political left/ % Voting Democrats | 26.0% | 33.9% | 0.50 |
| % Employed | 74.0% | 72.3% | −1.25 |
| % having parents with academic degrees | 32.0% | 33.9% | 0.28 |
| Age | 21.6 | 22.1 | −0.30 |
| Words found in puzzle | 12.1 | 12.8 | −0.31 |
| Observations | 100 | 112 |  |

Notes
The table presents summary statistics of the participants in the two experiments. Column 1 gives the summary statistics of the participants in the control treatment. Column 2 gives the summary statistics of the participants in the communal treatment. Column 3 gives the results of Wilcoxon rank sum test comparing the distributions. Workers retained is the average response to the question about how many workers the participants chose to retain. % Economics students is the % of students studying economics, accounting, business administration, banking and finance, or management. % women is the % of women. % married is the % of married participants. % political left/% voting democrats is the % of participants that vote for center-left/left-wing parties (Israeli data), or that vote for the Democratic party (U.S. data). % employed is the % of participants that work either part or full-time. Age is the average age of the participants. % having parents with academic degrees is the % of the participants that both their parents have academic degrees. Age is the participants' average. Words found in the puzzle are the average number of words that the participants found in the puzzles.

\* $p < 0.10$, \*\* $p < 0.05$, \*\*\* $p < 0.10$.